<div class="moz-text-flowed" style="font-family: -moz-fixed">
\documentclass[printer]{aa}
\usepackage{graphicx}
\usepackage{txfonts}
\usepackage{longtable}
\def\TiII{Ti{$\,$\sc II}}

\begin{document}
\title{Exploring interstellar titanium and deuterium abundances and 
other correlations
\footnote{Based on observations taken with the Ultraviolet and Visual 
Echelle Spectrograph (UVES) on the Very Large Telescope (VLT) Unit 2 
(Kueyen) at Paranal, Chile, operated by ESO.}}


    \author{R. Lallement
           \inst{1}
           \and
           G. H\'ebrard
           \inst{2}
           \and
           B.Y. Welsh
           \inst{3}
           }

    \offprints{R. Lallement}

    \institute{Service d'A{\'e}ronomie, UMR7620 CNRS, Universit\'e 
Versailles Saint-Quentin, BP3, F-91371 Verri{\`e}res-le-Buisson, 
France\\
               \email{rosine.lallement@aerov.jussieu.fr}
          \and
              Institut d'Astrophysique de Paris, UMR7095 CNRS, 
Universit\'e Pierre \& Marie Curie, 98$^{\rm bis}$ boulevard Arago,\\
           F-75014 Paris, France
              \email{hebrard@iap.fr}
          \and
             Space Sciences Laboratory, University of California, 7 
Gauss Way, Berkeley, CA 94720,USA\\
              \email{bwelsh@ssl.berkeley.edu}
Southern Observatory, Chile; proposal no. 77.C-0347A.
              }

    \date{Received ; accepted }


\abstract
{}
    {The origin of the observed variability of the gas-phase D/H ratio 
in the local interstellar medium is still debated, and in particular 
the role of deuterium depletion onto dust grains. Here we extend the 
study of the relationship between deuterium and  titanium, a 
refractory species and tracer of elemental depletion, and explore 
other relationships.}
  {We have acquired high resolution spectra for nine early-type stars 
using the VLT/UVES spectrograph, and detected the absorption lines of 
interstellar TiII.
  Using a weighted orthogonal distance regression (ODR) code and a 
special method to treat non symmetric errors, we compare the TiII
columns with the corresponding HI, DI and also OI columns. In 
parallel we perform the same comparisons for available FeII data.}
    {We find a significant correlation between TiII/HI and D/H in our 
data set, and, when combined with published results, we confirm and 
better constrain the previously established trends and extend the 
trends to low HI columns.

We exclude uncertainties in HI and OI columns as the main contributor 
to the derived metals-deuterium correlations by showing that the 
TiII/HI ratio is positively correlated with DI/OI. We find a similar 
correlation between FeII/HI and DI/OI.

The TiII gradients are similar or slightly smaller than for FeII, 
while one would expect larger variations on the basis of the higher 
condensation temperature of titanium. However we argue that 
ionisation effects introduce biases that affect iron and not titanium 
and may explain the gradient similarity.

We find a less significant negative correlation between the TiII/DI 
ratio and the hydrogen column, possibly a sign of different 
evaporation of D and metals according to the cloud properties. More 
TiII absorption data along very low H column lines-of-sight would be 
useful to improve the correlation statistics.
}
    {}


\titlerunning{Interstellar Titanium and Deuterium}
\authorrunning{Lallement et al.}
    \maketitle
%

\section{Introduction}
The measurement of the present-day ratio of deuterium (D) to hydrogen 
(H) in the Galaxy places important constraints on its chemical 
evolution and has been the goal of intense work, from the Copernicus 
era (Rogerson and York, 1973) to HST (eg. Linsky, 1998), IMAPS 
(Jenkins et al., 1999, Sonneborn et al., 2000) and FUSE (e.g. Moos et 
al., 2002).
Measurements inside the
Local Bubble (LB), the volume of tenuous gas surrounding the Sun 
(e.g. Snowden et al., 1998, Welsh et al., 1999, Lallement et al., 
2003) appear to be consistent
with a single value for D/H (Moos et al., 2002, H{\'e}brard \& Moos, 
2003, Oliveira et al., 2003), but other
measurements (e.g. Jenkins et al., 1999, Sonneborn et
al., 2000,  Hoopes et al., 2003, Friedman et al., 2006, H{\'e}brard 
et al., 2005, Oliveira \& H{\'e}brard, 2006) suggest variations of 
the interstellar D/H ratio
at larger distances, i.e. beyond $\simeq$ 150-200
parsecs. Quantitatively, within and at the periphery of the LB (log 
N(HI) $\leq$ 19.3 cm$^{-2}$), the
D/H ratio is of the order of 15 ppm (see Figure 1 of Linsky et al., 
2006), while lines-of-sight measurements with higher N(HI) show a 
strong variability, with maximum values of $\simeq$ 23 ppm. For a 
number of lines-of-sight (LOS) with log N(HI) $\leq$ 20.5 cm$^{-2}$, 
D/H is significantly lower, $\simeq$ 7 ppm.

According to Draine (2004) and Linsky et al. (2006) (see also Linsky 
2007), the wide range in the gas-phase D/H ratios is due to different 
amounts of deuterium depletion onto interstellar grains. Theoretical 
support for this interpretation is the preferential adsorption of 
deuterium onto interstellar grains, in particular PAHs (Jura, 1982, 
Draine, 2004). Observational support comes from the correlations 
between D/H and both Fe and Si depletions, as well as the correlations
with the rotational temperatures of H$_{2}$ for lines-of-sight 
possessing molecular gas (Linsky et al., 2006). The D/H trends quoted 
above may be explained by the depletion hypothesis if one 
hypothesizes a recent heating of the LB and
especially recent shocks in star-forming regions at its periphery. 
This interpretation has important consequences and implies that the 
total D/H ratio including deuterium in the gas and in the dust is at 
least 23 parts per million of hydrogen, which poses a challenge for 
models of galactic chemical evolution.

On the other hand, Oliveira \& H{\'e}brard (2006) show that high D/H 
values are also found along distant lines-of-sight with associated 
high interstellar columns. At high N(HI) the selection bias against 
high D/H values (due to dominance of dense, cold, and thus strongly 
depleted regions) should not lead to any variability (Oliveira et 
al., 2006). Moreover, D/O measurements do not fully support this 
depletion scenario, since the DI/OI ratio is found to be more 
constant than DI/HI (H\'ebrard et al., 2005). Interestingly the LB
D/H ratio derived from the D and O measurements is only of the order 
of 13 ppm (H{\'e}brard \& Moos 2003).
These authors suggest that the uncertainties in the measurement of HI 
in the data play an important role in the subsequently derived 
relationships. Two quotients with the same denominator remain 
naturally correlated if there are measurement errors on this 
denominator. FeII/HI  vs. DI/HI or FeII/OI vs. DI/OI relationships 
could result from this effect.

Infall of external unprocessed (or less processed) gas also creates 
D/H variability, and indeed a continuous infall is a general property 
in the Galaxy (Romano et al., 2006). Support for the existence of a 
significant infall into the Sun's neighborhood comes from D, O, N, 
and $^{3}$He abundance values  found in the local interstellar cloud 
that surrounds the Sun. Geiss et al. (2002)
interpret these abundances as being due to the local mixture of 
processed galactic gas and very weakly processed (Magellanic type) 
gas of intergalactic origin having fallen onto the disk. Such an 
infall could have occurred recently, leading to a present-day 
incomplete mixing. On the other hand, the wide D/H range implies a 
non realistic infall strength and a very weak mixing after the 
infall, as argued by Linsky et al. (2006).

It is therefore mandatory to investigate in more detail the depletion 
mechanisms and the associated deuterium behaviour. Prochaska et al. 
(2005) made an important step by showing a positive correlation 
between the
abundance of titanium, a refractory species, and that of D/H. Here we 
present results we have obtained in the southern hemisphere with
the Ultraviolet and Visual Echelle Spectrograph (UVES) at the ESO 
Very Large Telescope (VLT), and we add nine new TiII measurements 
towards D/H targets. In parallel to our observations and analysis, 
Ellison et al. (2007) also gathered and analysed new VLT-UVES  and 
Keck Observatory HIRES data. We have combined our results with their 
results and those from Prochaska et al. For targets in common, we 
have compared the two results. We have extended the correlative 
study, in particular to lower column-densities. We have used 
sophisticated methods to treat asymmetric errors in both correlated 
quantities.

Titanium is a particularly well suited element for the study of a 
potential correlation between depletion and deuterium abundance. It 
has a high depletion which is linked to its high condensation 
temperature. \TiII\ is optically accessible in the
near ultraviolet via the $\lambda\lambda$ 3229.199,3242.994,3383.768 
\AA\ absorption lines (wavelengths from Morton, 1991), and 
interstellar titanium has been observed in absorption since the 
1970's (e.g. Wallerstein and Goldsmith, 1974, Stokes, 1978, Hobbs, 
1984). Moreover, as already argued by these authors, \TiII\ is the 
dominant ionization state in HI regions due to the nearly exact 
coincidence of the ionization potentials of \TiII\ and HI.
Also, \TiII\ is coupled to HI via charge-exchange reactions (Field 
and Steigman, 1971). The observed close N(\TiII) vs. N(HI) 
relationship, in contrast with the more dispersed N(NaI)/N(HI) or 
N(CaII)/N(HI) relations (see e.g. Figure 7 of Welsh et al., 1997), 
results from those physical properties.

The detection of the interstellar TiII lines requires high spectral 
resolution (R $>$ 50,000)
  and appropriate optical systems and detectors adapted to the near 
ultraviolet wavelength range. Prochaska et al. (2005) used HIRES at 
the Keck Observatory and observed
absorption towards 7 northern hemisphere stars. Due to this limited 
number of targets, it is important to extend their study to other D/H 
targets. Moreover, the observed trend is strongly dependent on the 
measurement towards Feige 110, and indeed H\'ebrard et al. (2005) 
have shown that
the column of HI is probably significantly underestimated for this 
star. They also quote the contradicting case of $\gamma^{2}$Vel,  a 
line-of-sight with high D/H and low titanium abundance. Finally, 
while they covered a  range of logN(HI) from 19.85 and 21.05 
cm$^{-2}$, it is important to extend the study to lines-of-sight with 
lower HI columns, because there is a strong D/H variability for 
logN(H) $\leq$ 20 cm$^{-2}$ and no titanium column density data are 
currently available within this range.

Section 2 describes the observations, data analysis and TiII column 
determination. In section 3 we perform linear
comparison fits between TiII and other species, using our data in 
combination with other published data. We perform corresponding 
correlative studies for available FeII data. Section 4 discusses
the results.

\begin{figure*}
  \centering
  \includegraphics[width=16cm,angle=0]{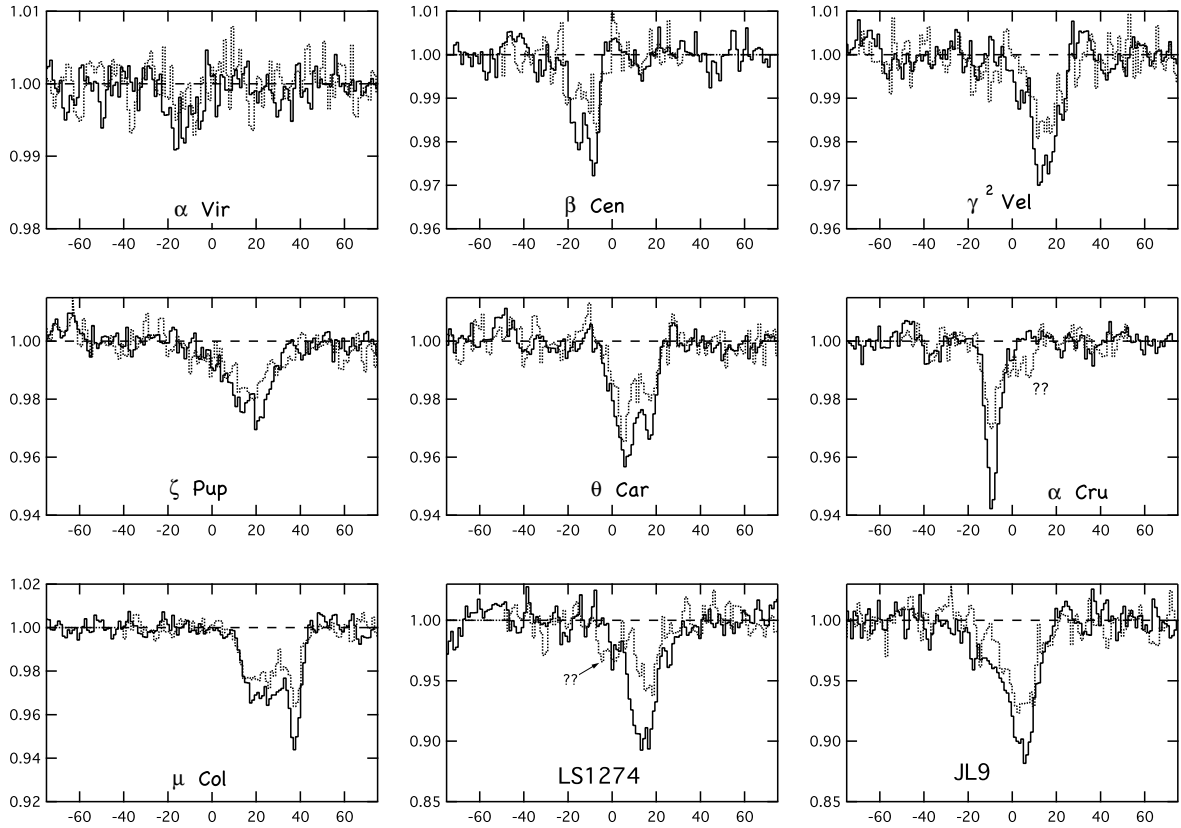}
    \caption{Normalised spectra around the TiII 3384 \AA\ (solid 
lines) and 3242 \AA\ (dashed lines) transitions for our target stars}. 
The abscissa is the heliocentric velocity in km/s.
          \label{spectra}
\end{figure*}

\section{Observations and absorption line analysis}
The spectroscopic data were obtained with the ESO VLT-UT2 telescope 
and the UVES spectrograph (Dekker et al., 2000) under the observing 
program 077.C-0547A.
We used the standard dichroic 346-580nm setting (DIC1) and the image 
slicer 1 which
  corresponds to R $\simeq$ 60,000 at 3400 \AA. This setup included 
the $\lambda\lambda$ 3229,3242,3384 \AA\ \TiII\ lines with the blue 
arm. The spectra were recorded in service mode.

The targets were seven bright early-type stars and two hot subdwarfs. 
They were selected from the list of D/H targets stars recently 
compiled by Linsky et al., (2006). Multiple short exposures were 
obtained for each of the bright stars and two long exposures for the 
subdwarfs.
Details of the observations are listed in Table 1.
The individual stellar spectra, processed through the UVES pipeline, 
have been co-added.

The spectral analysis and the correlative studies were
performed in the frame of the IGOR/WaveMetrics software, using
both built-in and specially developed routines.
In the present work we are interested in the total \TiII\ 
column-density values (a cloud-by-cloud analysis is beyond its 
scope), and since we are in the optically thin linear regime 
measurement of the total equivalent width is sufficient.
For this purpose, stellar continua around the strongest transitions 
$\lambda\lambda$ 3383.768, 3241.994 \AA\  were fit with a multi-order 
polynomial. The noise level was automatically calculated during the 
stellar continuum fit and assumed to prevail at the absorption line 
locations.  The resulting signal to noise ratios around the 3384\AA\ 
transition are given in Table 1. The normalised spectra corresponding 
to the two strongest transitions  are displayed in Figure 1. Note 
that these targets stars correspond to a two orders of magnitude 
variation of the HI column-density.

After division by the fitted continuum, equivalent widths are simply 
deduced from the normalised spectra. Errors on the total \TiII\ 
equivalent width are estimated  on the basis of the noise level and 
the number of pixels defining  the absorption.

Two of the target spectra ($\theta$ Car, $\zeta$ Pup) are affected by 
some instrumental interference fringes. For these stars the stellar 
continua were fit with the sum of  a multi-order polynomial and a 
sinusoidal function with an adjustable period, phase and amplitude. 
The correction is illustrated in Figure 2 for $\theta$ Car which 
corresponds to the maximum fringe amplitude. The difference in 
derived equivalent widths (or column densities) with and without the 
fringe correction is generally less than the error resulting from the 
noise.

The column-densities of TiII are deduced here from the strongest 3384 
\AA\ transition. On the basis of experimental data Pickering et al 
(2001) have revised the oscillator strength of this transition at 
f=0.358, some 5 percent higher than the Morton (1991) value f=0.3401. 
In order to be consistent with the previous analyses of Prochaska et 
al. and Ellison et al. we have used this revised value.
Note that the results of this work and previous ones on the existence 
and quality of correlations remain unchanged if this oscillator 
strength is revised again. Only resulting gradients would have to be 
scaled according to the changes.

For the $\lambda\lambda$ 3242\AA\ transition we first used the Morton 
(1991) atomic data, then the revised values quoted by Prochaska et al 
(2005). Their atomic data obtained from Morton (2003, private 
communication) are significantly different from the older ones, with 
the oscillator strength increased from f=0.183 to 0.232.
We could check from the simultaneous fits of the  two lines that 
those revised values are in better agreement with data than the 
Morton (1991) values and we show only the second results.

We have also performed line-fitting, using the program described in 
Sfeir et al. (1999), i.e., the residual intensity absorption profiles 
were fit to combinations of Voigt profiles, convolved by the UVES 
instrumental function.   We have limited the temperature to the 
100-15,000K  range, which corresponds to realistic values in the 
local ISM, and we have simply chosen the minimum number of clouds 
allowing a good visual  adjustment to the data. In parallel we 
performed a profile fitting analysis with the Owens.f software of 
Lemoine et al. (2002) for one of the targets and the results are 
consistent.
The cloud-by-cloud structure in those line-fitting results should not 
be considered as the best solutions, since they should also be 
compared in future work with ithe structure derived from other 
elements. However, they provide a consistency check.
Simultaneous fitting of the two transitions are shown in Figures 3 
and 4. We have not included $\alpha$Vir, $\beta$Cen and $\alpha$Cru, 
for which the column is small and the detection of the weakest line 
is of insufficient quality and does not add to the 3384\AA\ detection.

Although we do not specifically use the column densities or the 
cloud-by-cloud structure resulting from this exercise,
fitting the two 3384 and 3242 \AA\ lines,as said above, has provided 
a consistency check: Lines other than TiII and (or) non-random noise 
features are revealed by this procedure.

While performing this exercise we noticed the following departures 
from the model absorptions.
For LS1274 the $\lambda\lambda$ 3241.994 \AA\ region is affected by 
an additional absorption that has no counterpart at 3384\AA\, not at 
3229 \AA\ (the absorption is strong enough for this star to be 
detectable at this wavelength). Looking at the entire spectrum, we 
found that at variance with the other targets and especially the 
second subdwarf JL9, the spectrum is characterized by a very large 
number of narrow features. We could not determine whether these 
narrow lines have a circumstellar or stellar origin (LS1274 is cooler 
than JL9) origin, and their study is beyond the scope of this paper. 
For this star we have calculated the column-density without including 
this additional absorption that is not seen in the stronger 
transition and is indicated in Figure 1. A second, although smaller, 
anomaly is found for $\alpha$Cru, for which there is an absorption 
feature seen at 3342 \AA\ and not for the strongest transition. The 
noise level however is such that a non-random noise feature may be 
responsible.
After examination of the two transitions and of these line fit 
results we have accordingly increased the error bars on the 
equivalent widths and resulting columns to take into account such non 
random noise features.

We also have searched for a possible contamination by a stellar line 
by shifting each spectrum to the stellar reference frame and 
overplotting the resulting shifted spectra. No particular absorption 
at a given wavelength is revealed by such a method. The two subdwarfs 
however are excluded because stellar radial velocities for those 
targets are uncertain.

For all targets except $\mu$Col, the equivalent width ratios are 
compatible with the theoretical value of $\simeq$1.65 deduced from 
the revised atomic data, when taking into account noise features and 
the LS1274 blend, and it is possible to obtain a reasonable 
simultaneous fit to the two transitions. Although in the case of 
$\mu$Col the spectrum is of excellent quality, and the two 
transitions clearly show the same structure, we could not fit them 
simultaneously in a satisfying way (see Figure 3). More specifically, 
the narrow line in the red is correctly fitted, but the broad feature 
in the blue is not, and the probability of this being due to random 
noise is less than 5\% (2$\sigma$ equivalent width deficit). We 
suspect a non-interstellar origin for this broad absorption, and will 
discuss the peculiarity of this star later.

We have also verified that for all our targets neutral titanium (TiI) 
is negligible with respect to \TiII\ (see section 1). For that 
purpose we checked for the presence of $\lambda\lambda$ 3342,3636 
\AA\  TiI absorption lines within the velocity interval deduced from 
the observed \TiII\ lines. We could not detect any line absorption. 
Since oscillator strengths for those TiI transitions are similar to 
the \TiII\  ones, this implies that the contribution from TiI is 
negligible.

\begin{figure}
    \centering
   \includegraphics[width=8cm]{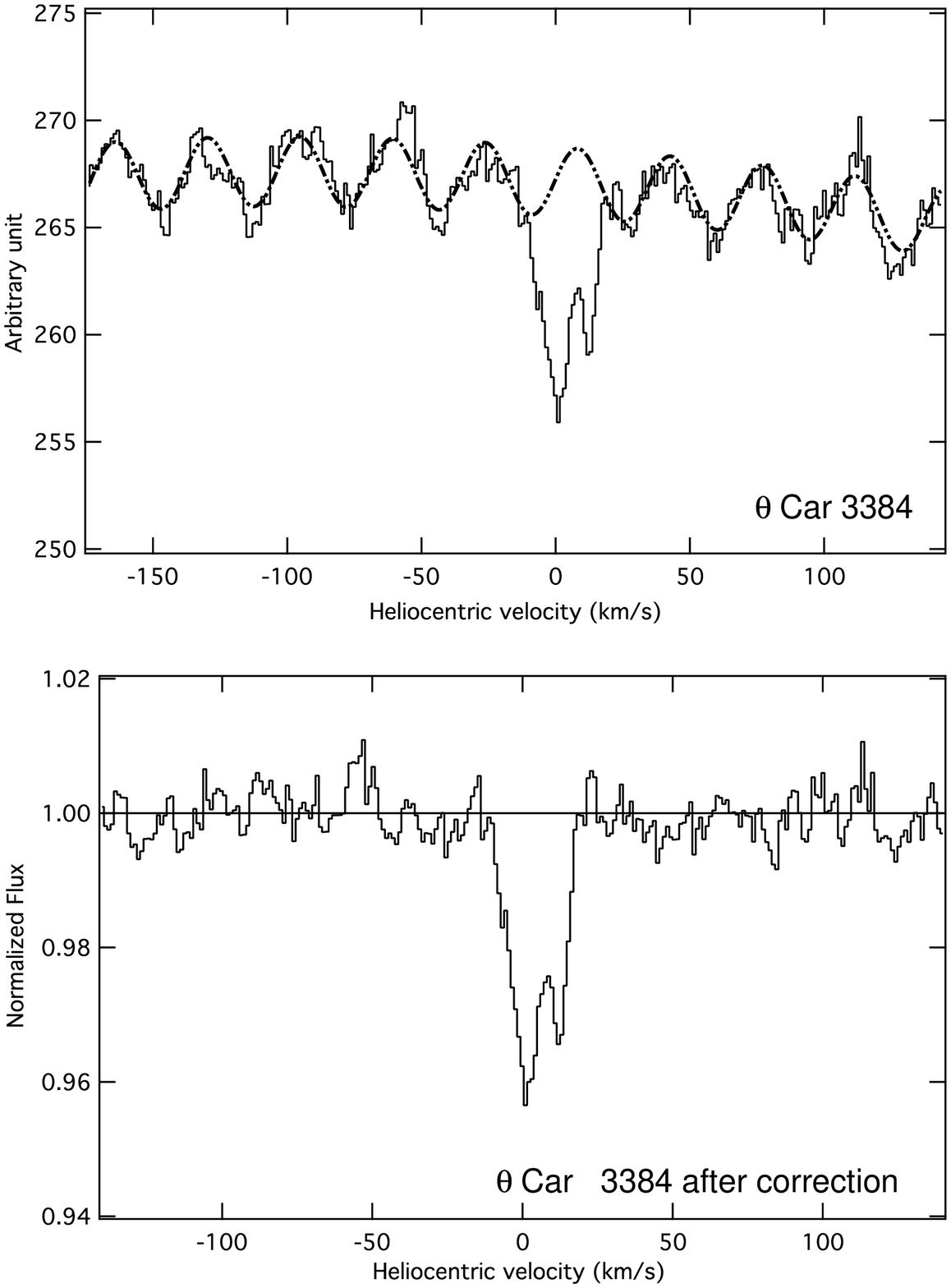 }
       \caption{Illustration of fringe correction to the data: the 
extreme case of $\theta$Car. }
          \label{tetacar}
\end{figure}

\begin{figure*}
    \centering
   \includegraphics[width=14cm]{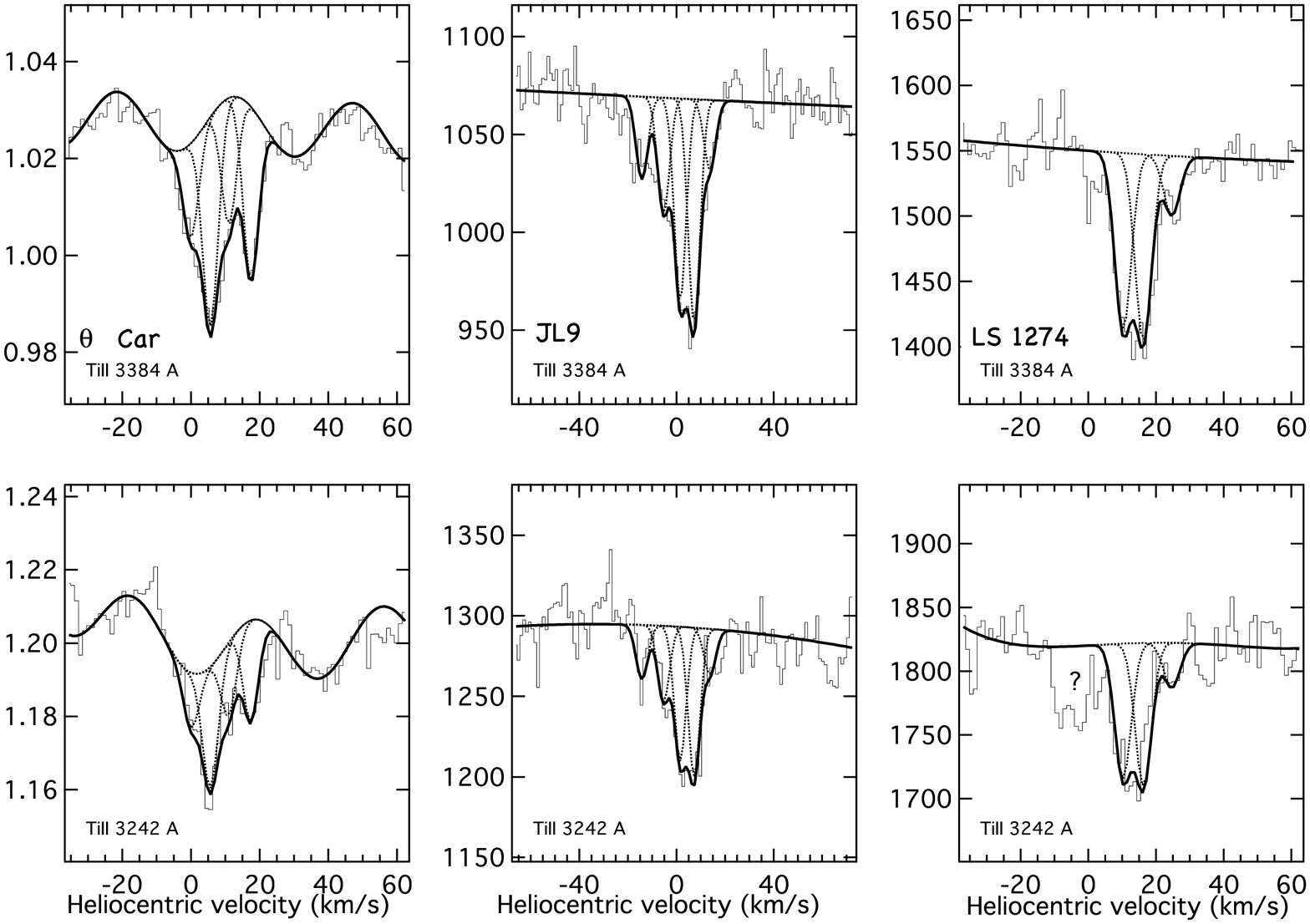 }
       \caption{3394-3242\AA\ simultaneous line-fitting used as a 
consistency checked: $\theta$Car,JL9, and LS1274. The blue-shifted 
absorption at 3384\AA\ in LS1274 has no counterpart at 3242\AA\ and 
is very likely stellar (see text).}
          \label{gamvel}
\end{figure*}

\begin{figure*}
    \centering
   \includegraphics[width=14cm]{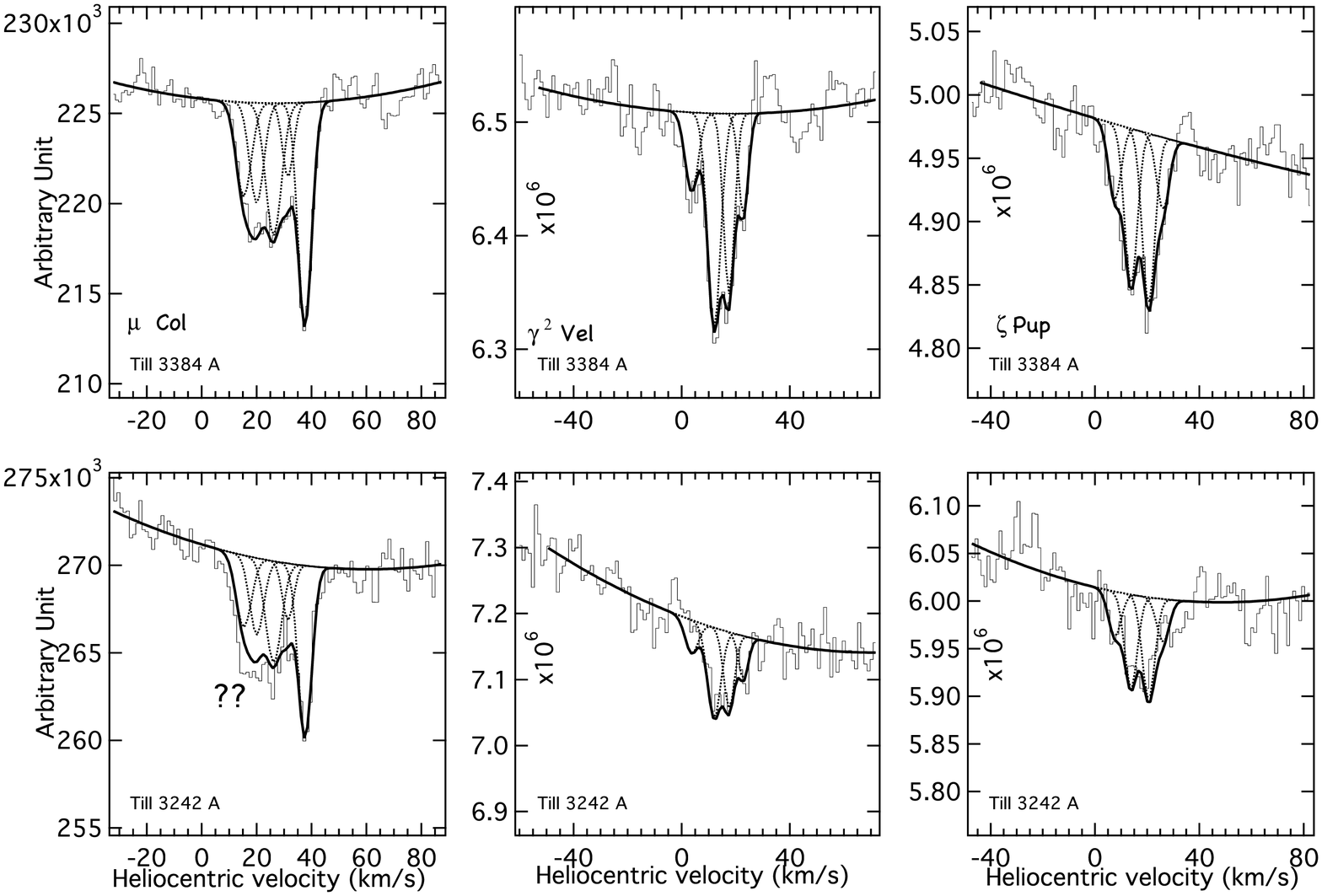 }
       \caption{Same as Figure 3: $\gamma^{2}$Vel,$\zeta$ Pup, and 
$\mu$ Col. The broad feature in $\mu$ Col cannot be fitted 
satisfyingly (see text)}
          \label{gamvel}
\end{figure*}

\begin{table*}
\caption{Summary of observations and titanium results. (1) distance 
recently revised by Millour et al (2007). *: S/N after fringe 
correction}
\label{table:1}
\centering
\begin{tabular}{c c c c l l l l l }     
\hline\hline
Target & $lII$ & $bII$ & d(pc) & type &  Obs-Date  & Exp-time(s) & SN 
& LogN(TiII) \\
\hline
    $\alpha$ Vir & 316.1 & 50.8 & 80$\pm$6 & B1III-IV & 22-05-06 & 
50x.35 & 380 & 10.49 $\pm$ 0.13  \\
    $\beta$ Cen & 311.8 & 1.2    & 161$\pm$15 & B1III & 15-04-06 & 
50x0.2 & 340 & 10.99 $\pm$ 0.04 \\
    $\alpha$ Cru & 300.1 & -0.36 & 98$\pm$6 & B1  & 10-04-06 & 50x0.15 
& 290 & 11.12 $\pm$ 0.035\\
    $\gamma^{2}$ Vel & 262.8 & $-7.7$  & 368(+38-13)\footnote{recently 
revised by Millour et al., 2007} & WR+O & 09-04-06& 50x0.6 & 320 & 
11.11 $\pm$ 0.035\\
    $\mu$ Col & 237.2 & $-27.1$    & 396$\pm$100 & O9.5 & 01-04-06 & 
40x8 & 310 & 11.50 $\pm$ 0.02\\
$\zeta$ Pup & 256.0 & -4.7   & 429$\pm$94 & O5I & 29-03-06 & 45x0.7 & 
250* & 11.16 $\pm$ 0.06\\
$\theta$ Car & 289.6 & $-4.9$   & 135$\pm$9 & B0V & 10-04-06 & 40x0.9 
& 320* & 11.31 $\pm$ 0.023\\
JL9 & 322.6 & -27.0    & 590$\pm$160  & sdO & 19-05-06 & 2x2700 & 280 
& 11.88 $\pm$ 0.10 \\
LS 1274 & 277.0 & -5.3    & 580$\pm$100 & sdB & 10-04-06  & 2x2100 & 
110 & 11.68 $\pm$0.04 \\

\hline\hline
\end{tabular}
\end{table*}

\begin{table*}
\caption{Compilation of D/H, D/O, titanium and iron data. Errors bars 
are quoted in the usual way with negative errors as indices and 
positive errors as exponants. References of TiII data used in the 
correlations are: (a) Prochaska et al.,2005, (b), Ellison et al., 
2007, (c), this work. Targets in common with Ellison et al (2007) are 
marked with an asterix. The column-densities derived by these authors 
are log(TiII)= 11.10$\pm$0.02; 11.31$\pm$0.01; 11.12$\pm$0.02; 
11.05$\pm$0.04; 11.07$\pm$0.04 for $\gamma^{2}$ Vel; $\theta$ Car; 
$\alpha$ Cru; $\beta$ Cen and $\zeta$ Pup respectively and are in 
agreement with our determinations within uncertainties.
D, H, D/H, D(Fe) are the values compiled by Linsky et al., (2006), 
while D, O, D/O are those compiled by Oliveira et al., (2006). The 
three targets with a very high D/O and discussed in sect. 3.2 are 
marked with a \#; the first D/O value is the published value used in 
Figure 9a. The second number is the reduced value used in Fig. 9a 
(see section 3.2).}
\label{table2}
\centering
\begin{tabular}{c c c c l l l l c }     
\hline\hline
Target & log(NHI) & log(DI) & D/H(ppm) & DI/OI x10$^{2}$ & D(Fe) & 
log(TiII)  & TiII/HI x10$^{10}$  & ref\\
\hline

SiriusB & 17.6 $^{ 0.14 }_{ 0.12 }$& 12.81 $^{ 0.09 }_{ 0.09 }$& 16.2 
$^{ 7.2 }_{ 6.4 }$& 3.9 $^{ 0.8 }_{ 0.8 }$& -1.11 $^{ 0.12 }_{ 0.14 
}$&      &   $^{   }_{   }$& \\
36Oph & 17.85 $^{ 0.075 }_{ 0.075 }$& 13.025 $^{ 0.01 }_{ 0.01 }$& 15 
$^{ 2.5 }_{ 2.5 }$&   $^{   }_{   }$& -0.65 $^{ 0.27 }_{ 0.27 }$& 
&   $^{   }_{   }$&  \\
$\epsilon$ Eri & 17.875 $^{ 0.035 }_{ 0.035 }$& 13.03 $^{ 0.03 }_{ 
0.03 }$& 14.3 $^{ 1.6 }_{ 1.6 }$&   $^{}_{   }$& -1.07 $^{ 0.1 }_{ 
0.1 }$&      &   $^{   }_{   }$&  \\
31 Com & 17.884 $^{ 0.03 }_{ 0.03 }$& 13.19 $^{ 0.025 }_{ 0.025 }$& 
20.2 $^{ 1.9 }_{ 1.9 }$&   $^{   }_{   }$& -0.78 $^{ 0.1 }_{ 0.1 }$& 
&   $^{   }_{   }$&  \\
HZ43 & 17.93 $^{ 0.03 }_{ 0.03 }$& 13.15 $^{ 0.02 }_{ 0.02 }$& 16.6 
$^{ 1.4 }_{ 1.4 }$& 4.6 $^{ 0.5 }_{ 0.5 }$& -1.21 $^{ 0.04 }_{ 0.04 
}$&      &   $^{   }_{   }$&  \\
Procyon & 18.06 $^{ 0.05 }_{ 0.05 }$& 13.26 $^{ 0.027 }_{ 0.027 }$& 
16 $^{ 2 }_{ 2 }$&   $^{   }_{   }$& -1.24 $^{ 0.05 }_{ 0.05 }$& 
&   $^{   }_{   }$&  \\
$\beta$ Cas & 18.132 $^{ 0.025 }_{ 0.025 }$& 13.36 $^{ 0.03 }_{ 0.03 
}$& 16.9 $^{ 1.6 }_{ 1.6 }$&   $^{   }_{   }$& -1.22 $^{ 0.1 }_{ 0.1 
}$&      &   $^{   }_{   }$&  \\
G191-B2B & 18.18 $^{ 0.09 }_{ 0.09 }$& 13.4 $^{ 0.035 }_{ 0.035 }$& 
16.6 $^{ 4.1 }_{ 4.1 }$& 3.5 $^{ 0.4 }_{ 0.4 }$& -0.58 $^{ 0.09 }_{ 
0.09 }$&      &   $^{   }_{   }$&  \\
$\beta$ Gem & 18.261 $^{ 0.037 }_{ 0.037 }$& 13.43 $^{ 0.05 }_{ 0.05 
}$& 14.8 $^{ 2.2 }_{ 2.2 }$&   $^{   }_{   }$& -1.29 $^{ 0.05 }_{ 
0.05 }$&      &   $^{   }_{   }$&  \\
Capella & 18.239 $^{ 0.035 }_{ 0.035 }$& 13.44 $^{ 0.02 }_{ 0.02 }$& 
15.9 $^{ 1.5 }_{ 1.5 }$& 2.6 $^{ 1.2 }_{ 1.2 }$& -1.2 $^{ 0.04 }_{ 
0.04 }$&      &   $^{   }_{   }$&  \\
$\alpha$ Tri & 18.327 $^{ 0.035 }_{ 0.035 }$& 13.45 $^{ 0.05 }_{ 0.05 
}$& 13.3 $^{ 2 }_{ 2 }$&   $^{   }_{   }$& -1.13 $^{ 0.2 }_{ 0.2 }$& 
&   $^{   }_{   }$&  \\
WD0621 376 & 18.7 $^{ 0.15 }_{ 0.15 }$& 13.85 $^{ 0.045 }_{ 0.045 }$& 
14.1 $^{ 6 }_{ 6 }$& 3.9 $^{ 0.6 }_{ 0.6 }$& -1.21 $^{ 0.2 }_{ 0.2 
}$&     &   $^{   }_{   }$&  \\
WD2211 495 & 18.76 $^{ 0.15 }_{ 0.15 }$& 13.94 $^{ 0.05 }_{ 0.05 }$& 
15.1 $^{ 6.5 }_{ 6.5 }$& 4 $^{ 0.6 }_{ 0.6 }$& -0.96 $^{ 0.16 }_{ 
0.16 }$&      &   $^{   }_{   }$&  \\
WD1634 573 & 18.85 $^{ 0.06 }_{ 0.06 }$& 14.05 $^{ 0.025 }_{ 0.025 
}$& 15.8 $^{ 2.5 }_{ 2.5 }$& 3.5 $^{ 0.3 }_{ 0.3 }$& -1.22 $^{ 0.09 
}_{ 0.09 }$&      &   $^{   }_{   }$&  \\
$\alpha$ Vir & 19 $^{ 0.1 }_{ 0.1 }$& 14.2 $^{ 0.2 }_{ 0.1 }$& 15.8 
$^{ 10.8 }_{ 5.8 }$& 4.2 $^{ 2 }_{ 2 }$& -1.29 $^{ 0.1 }_{ 0.12 }$& 
10.49 $\pm$ 0.13 & 32.4 $^{ 14.4 }_{ 10 }$& c \\
GD246 & 19.11 $^{ 0.025 }_{ 0.025 }$& 14.29 $^{ 0.045 }_{ 0.045 }$& 
15.1 $^{ 1.9 }_{ 1.9 }$& 4.2 $^{ 0.6 }_{ 0.6 }$& -1.26 $^{ 0.06 }_{ 
0.06 }$&      &   $^{   }_{   }$&  \\
$\lambda$ Sco & 19.23 $^{ 0.03 }_{ 0.03 }$& 14.11 $^{ 0.09 }_{ 0.07 
}$& 7.6 $^{ 1.4 }_{ 1.8 }$& 1.8 $^{ 0.4 }_{ 0.3 }$& -1.64 $^{ 0.05 
}_{ 0.05 }$&      &   $^{   }_{   }$&  \\
$\beta$ Cen & 19.63 $^{ 0.1 }_{ 0.1 }$& 14.7 $^{ 0.2 }_{ 0.2 }$& 11.7 
$^{ 7.5 }_{ 7.5 }$&   $^{   }_{   }$& -1.16 $^{ 0.11 }_{ 0.11 }$& 
10.99 $\pm$ 0.04 & 24 $^{ 6.9 }_{ 5.4 }$& *c \\
$\gamma^{2}$ Vel & 19.71 $^{ 0.026 }_{ 0.026 }$& 15.05 $^{ 0.03 }_{ 
0.03 }$& 21.8 $^{ 2.1 }_{ 2.1 }$&   $^{   }_{   }$& -1.21 $^{ 0.11 
}_{ 0.11 }$& 11.11 $\pm$ 0.035 & 26.3 $^{ 2.5 }_{ 2.3 }$&*c \\
BD284211 & 19.85 $^{ 0.02 }_{ 0.02 }$& 14.99 $^{ 0.025 }_{ 0.025 }$& 
13.9 $^{ 1 }_{ 1 }$& 4.7 $^{ 0.4 }_{ 0.4 }$& -1.2 $^{ 0.1 }_{ 0.1 }$& 
11.08 $\pm$ 0.08 & 17 $^{ 3.4 }_{ 2.9 }$& a \\
$\alpha$ Cru & 19.85 $^{ 0.07 }_{ 0.1 }$& 14.95 $^{ 0.05 }_{ 0.05 }$& 
12.6 $^{ 3.6 }_{ 2.7 }$&   $^{   }_{   }$& -1.3 $^{ 0.14 }_{ 0.12 }$& 
11.12 $\pm$ 0.035 & 19.5 $^{ 3.9 }_{ 3.3 }$&*c \\
$\mu$ Col & 19.86 $^{ 0.015 }_{ 0.015 }$& 14.7 $^{ 0.3 }_{ 0.1 }$& 
6.9 $^{ 6.9 }_{ 1.8 }$&   $^{   }_{   }$& -1.18 $^{ 0.02 }_{ 0.02 }$& 
11.50 $\pm$ 0.02 & 45.7 $^{ 3.3 }_{ 3.1 }$& c \\
Lan 23 & 19.89 $^{ 0.25 }_{ 0.04 }$& 15.23 $^{ 0.065 }_{ 0.065 }$& 
21.9 $^{ 17.4 }_{ 4.1 }$& 3.2 $^{ 1.6 }_{ 1.6 }$& -1.31 $^{ 0.07 }_{ 
0.25 }$&      &   $^{   }_{   }$&  \\
$\zeta$ Pup & 19.96 $^{ 0.026 }_{ 0.026 }$& 15.11 $^{ 0.06 }_{ 0.06 
}$& 14 $^{ 2.3 }_{ 2.3 }$&   $^{   }_{   }$& -1.28 $^{ 0.17 }_{ 0.07 
}$& 11.16 $\pm$ 0.06 & 16.6 $^{ 2.9 }_{ 2.5 }$& *c \\
TD1 32709 & 20.03 $^{ 0.1 }_{ 0.1 }$& 15.3 $^{ 0.05 }_{ 0.05 }$& 18.6 
$^{ 5.3 }_{ 5.3 }$& 7.59;3.75 $^{ 2.17 }_{ 1.76 }$& -1.53 $^{ 0.14 
}_{ 0.14 }$&      &   $^{   }_{   }$&\#  \\
WD1034+001 & 20.07 $^{ 0.07 }_{ 0.07 }$& 15.4 $^{ 0.07 }_{ 0.07 }$& 
21.4 $^{ 5.3 }_{ 4.5 }$& 6.31;3.15 $^{ 1.79 }_{ 1.38 }$& -1.42 $^{ 
0.12 }_{ 0.12 }$& 11.1 $\pm$ 0.2 & 10.7 $^{ 6.7 }_{ 4.1 }$& b\# \\
BD+39 3226 & 20.08 $^{ 0.09 }_{ 0.09 }$& 15.15 $^{ 0.05 }_{ 0.05 }$& 
11.7 $^{ 3.1 }_{ 3.1 }$& 5.62;2.81 $^{ 1.61 }_{ 1.31 }$& -1.38 $^{ 
0.11 }_{ 0.11 }$& 11.49 $\pm$ 0.08 &   $^{   }_{   }$& b\# \\
Feige110 & 20.14 $^{ 0.065 }_{ 0.1 }$& 15.47 $^{ 0.03 }_{ 0.03 }$& 
21.38 $^{ 5.7 }_{ 3.8 }$& 2.6 $^{ 0.5 }_{ 0.5 }$&  $^{   }_{   }$& 
11.59 $\pm$ 0.03 & 28.2 $^{ 4.9 }_{ 4.2 }$& a \\
$\iota$ Ori & 20.16 $^{ 0.06 }_{ 0.07 }$& 15.3 $^{ 0.04 }_{ 0.05 }$& 
14.13 $^{ 2.8 }_{ 2.8 }$& 3.5 $^{ 0.8 }_{ 0.8 }$& -1.41 $^{ 0.21 }_{ 
0.16 }$& 11.31 $\pm$ 0.03 & 14.1 $^{ 2.5 }_{ 2.1 }$& a \\
$\gamma$ Cas & 20.16 $^{ 0.08 }_{ 0.1 }$& 15.15 $^{ 0.04 }_{ 0.05 }$& 
9.8 $^{ 2.3 }_{ 2.7 }$& 2.5 $^{ 0.4 }_{ 0.4 }$& -1.53 $^{ 0.14 }_{ 
0.18 }$&      &   $^{   }_{   }$&  \\
$\delta$ Ori & 20.19 $^{ 0.03 }_{ 0.03 }$& 15.06 $^{ 0.06 }_{ 0.04 
}$& 7.4 $^{ 1.2 }_{ 0.9 }$& 2.5 $^{ 0.4 }_{ 0.4 }$& -1.56 $^{ 0.04 
}_{ 0.04 }$& 11.15 $\pm$ 0.04 & 9.1 $^{ 1.1 }_{ 1 }$& a \\
$\theta$ Car & 20.28 $^{ 0.1 }_{ 0.1 }$& 14.98 $^{ 0.18 }_{ 0.21 }$& 
5 $^{ 2.9 }_{ 3.4 }$&   $^{   }_{   }$& -1.54 $^{ 0.11 }_{ 0.11 }$& 
11.31 $\pm$ 0.023 & 11.2 $^{ 2.9 }_{ 2.3 }$& *c \\
$\epsilon$ Ori & 20.4 $^{ 0.08 }_{ 0.1 }$& 15.25 $^{ 0.05 }_{ 0.05 
}$& 6.3 $^{ 1.8 }_{ 1.5 }$& 1.9 $^{ 0.3 }_{ 0.3 }$& -1.7 $^{ 0.14 }_{ 
0.12 }$& 11.4 $\pm$ 0.03 & 10.0 $^{ 2.3 }_{ 1.9 }$& a \\
PG 0038+199 & 20.48 $^{ 0.04 }_{ 0.04 }$& 15.76 $^{ 0.04 }_{ 0.04 }$& 
19.1 $^{ 2.6 }_{ 2.6 }$& 2.4 $^{ 1 }_{ 0.5 }$& -1.51 $^{ 0.05 }_{ 
0.04 }$&      &   $^{   }_{   }$&  \\
JL9 & 20.78 $^{ 0.05 }_{ 0.05 }$& 15.78 $^{ 0.06 }_{ 0.06 }$& 10 $^{ 
1.9 }_{ 1.9 }$& 1.9 $^{ 0.8 }_{ 0.8 }$& -1.54 $^{ 0.1 }_{ 0.1 }$& 
11.88 $\pm$ 0.1 & 13.2 $^{ 3.8 }_{ 2.9 }$& c \\
HD195965 & 20.95 $^{ 0.025 }_{ 0.025 }$& 15.88 $^{ 0.07 }_{ 0.07 }$& 
8.51 $^{ 1.6 }_{ 1.6 }$& 1.3 $^{ 0.3 }_{ 0.3 }$& -1.59 $^{ 0.03 }_{ 
0.03 }$& 12.02 $\pm$ 0.02 & 11.7 $^{ 0.8 }_{ 0.8 }$& a \\
LS1274 & 20.98 $^{ 0.04 }_{ 0.04 }$& 15.86 $^{ 0.09 }_{ 0.09 }$& 7.6 
$^{ 1.9 }_{ 1.9 }$& 1.8 $^{ 0.5 }_{ 0.5 }$& -1.62 $^{ 0.08 }_{ 0.08 
}$& 11.68 $\pm$ 0.04 & 5.2 $^{ 0.8 }_{ 0.7 }$& c \\
HD191877 & 21.05 $^{ 0.05 }_{ 0.05 }$& 15.94 $^{ 0.11 }_{ 0.06 }$& 
7.8 $^{ 2.4 }_{ 1.5 }$& 2.5 $^{ 1 }_{ 1 }$& -1.55 $^{ 0.05 }_{ 0.05 
}$& 12.24 $\pm$ 0.02 & 15.5 $^{ 1.9 }_{ 1.7 }$& a \\
HD41161 & 21.08 $^{ 0.09 }_{ 0.09 }$& 16.41 $^{ 0.05 }_{ 0.05 }$& 
21.4 $^{ 5.7 }_{ 4.5 }$&   $^{   }_{   }$&  $^{   }_{   }$& 12.28 
$\pm$ 0.04 & 15.8 $^{ 4.1 }_{ 3.3 }$& b \\
HD53975 & 21.14 $^{ 0.06 }_{ 0.06 }$& 16.15 $^{ 0.07 }_{ 0.07 }$& 
10.2 $^{ 2.4 }_{ 2 }$&   $^{   }_{   }$&  $^{   }_{   }$& 12.13 $\pm$ 
0.04 & 9.8 $^{ 1.7 }_{ 1.5 }$& b \\
HD 90087 & 21.22 $^{ 0.05 }_{ 0.05 }$& 16.16 $^{ 0.06 }_{ 0.06 }$& 
8.7 $^{ 1.7 }_{ 1.7 }$& 1.7 $^{ 0.4 }_{ 0.4 }$& -1.45 $^{ 0.05 }_{ 
0.05 }$&      &   $^{   }_{   }$&  \\

\hline\hline
\end{tabular}
\end{table*}

\begin{figure*}
\centering
\includegraphics[width=15cm]{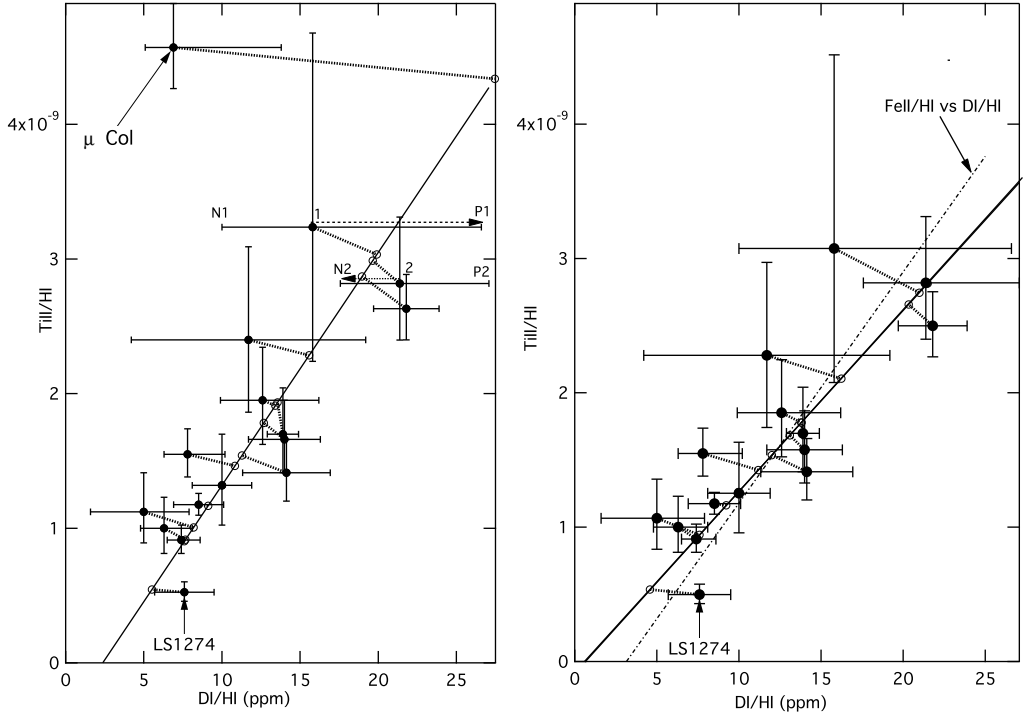}
       \caption{Ionized titanium to neutral hydrogen column ratio  vs. 
DI/HI with (left) and without (right) the star $\mu$Col. The method 
used to take into account non symmetric error bars for the abscissa 
is illustrated in the left figure. For the data point 1 (resp. 2), 
because it is located on the left (resp. right) side of the fitted 
line, the selected error bar is P1 (resp. N2). The same method is 
applied for the errors in the ordinate parameter. The corresponding 
linear fit for FeII is shown as a dotted line in the right figure.}
          \label{tetacar}
\end{figure*}

\begin{figure}
    \centering
   \includegraphics[width=8cm]{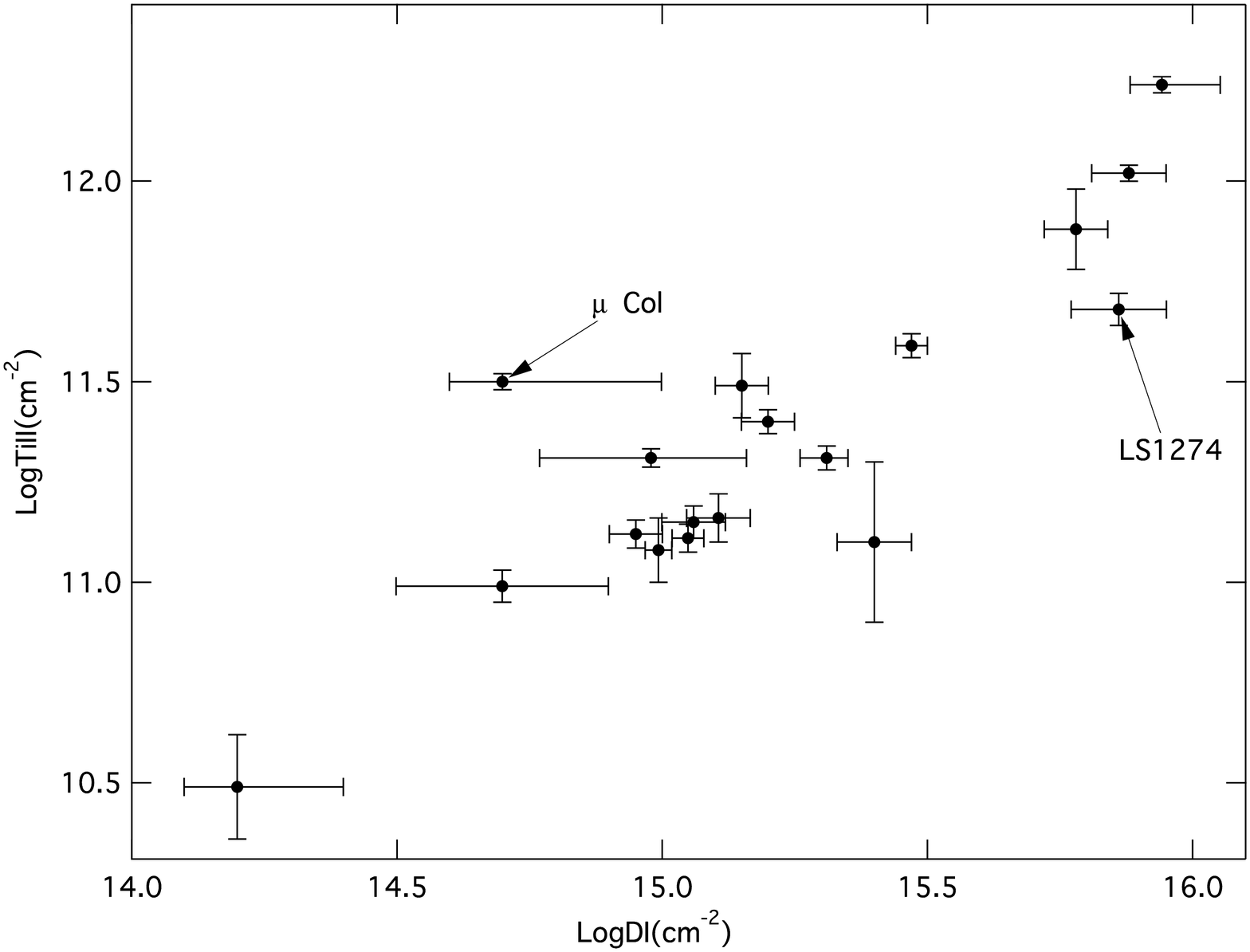}
       \caption{Titanium vs. deuterium columns. The star $\mu$Col is 
clearly an "outlier".}
          \label{FigTivsD}
\end{figure}

\begin{figure}
\centering
\includegraphics[width=8cm]{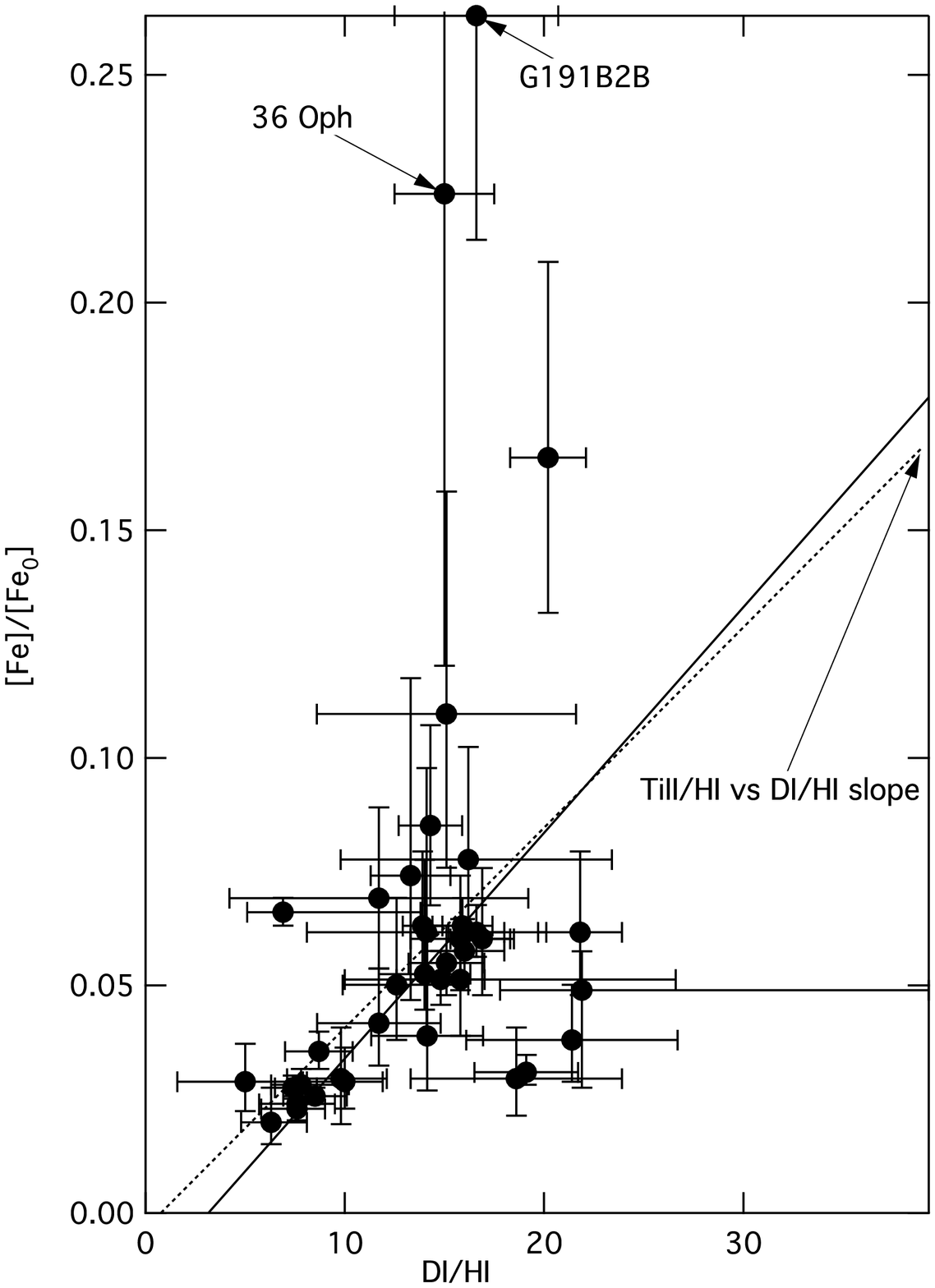}
       \caption{Ionized iron abundance normalised to 
(Fe/H)$_{0}$=-4.55 (Asplund et al., 2005)  vs. DI/HI. This figure is 
identical to Figure 3 from Linsky et al.(2006) except for the  linear 
scale. The best fit linear correlation using the Orthogonal Distance 
Regression (ODR) is shown as a solid line, and for comparison the 
titanium gradient obtained with the same method (Figure 5) is shown 
as a dashed line after the appropriate scaling to obtain a similar 
ordinate for D/H=20. It can be seen that the dispersion is 
significantly larger for FeII than for TiII. Lines-of-sight filled 
with strongly ionized gas (G191-B2B, 36 Oph) are positive extreme 
"outliers", while those corresponding to very large columns tend to 
produce negative "outliers".}
          \label{linfe-dh}
\end{figure}

%

\section{Comparisons between TiII,DI,HI,OI,FeII}

\subsection{Methods and selected targets}
In the following sections, we have combined our titanium data (9 
targets covering logN(HI) from 19.0 to 20.98)  with those of 
Prochaska et al. (7 stars with logN(HI) from 19.85 to 21.05). For all 
abundance correlations we have adopted the D/H ratio, logN(HI), N(DI) 
and iron depletion, D(Fe), compiled in Table 1 of Linsky et al. 
(2006). In addition, we have used the DI/OI ratios compiled in Table 
12 of Oliveira et al. (2006). All the data and errors used in this 
article are listed in Table 2.
For all correlations we have made use of the maximum amount of 
published data. Because OI and FeII measurements are not available 
for exactly the same targets, the correlation fits use different 
target lists.

After completion of this work, Ellison et al. (2007) made available 
results from  new
Keck and UVES TiII measurements. Some of their targets are already 
included in our new list. For those targets we have used our spectral 
results, resulting columns and errors. For those targets in common we 
also checked for consistency of the two measurements and found that 
our column density intervals overlap (see the caption of Table 2). 
There are 4 additional stars in Ellison et al. that were not included 
in our study. We have added these 4 targets in Table 2, and we have 
performed the same correlation fits with and without those four 
targets based on the correlation analysis method
discussed below. Four of our measurements are not part of the Ellison 
et al. study, the low column target $\alpha$Vir, the two faint 
subdwarfs, and $\mu$Col, for which Ellison et al. had an upper limit 
only.

We have also compared our column density results with previous 
measurements and analyses. We have disagreements with the results of 
Welsh et al. (1997).

While for  $\mu$Col  we differ by 10\% only, in the case of 
$\zeta$Pup our column log(TiII)= 11.18$\pm$ 0.06 is a factor of 2 
below their quoted value log(TiII)= 11.52$\pm$ 0.06. On the other 
hand their upper limits of log(TiII)=10.79 and 10.97 for 
$\gamma^{2}$Vel and  $\theta$Car respectively are a factor of two 
below our measured values of log(TiII)=11.13$\pm$0.035 and 
10.97$\pm$0.023.
We note that their NaI and CaII measurements are in excellent 
agreement with other authors (e.g. Welty, Morton
and Hobbs 1996), and the discrepancy in the TiII results could be due 
to an incomplete removal of the instrumental
background signal which is particularly difficult to measure at these 
very short wavelengths.
The D/H ratio depends on the amount of molecular hydrogen. For most 
of the targets here the relative amount of molecular hydrogen is 
small and we assume N(H)=N(HI) as well as D/H= DI/HI. For three 
targets only ($\theta$ Car, HD90087 and PG0038+199 ) that have 
significant molecular contribution we use the numbers from Table 2 of 
Linsky et al. (2006) and use N(H)=N(HI)+2N(H2) and D/H= 
(N(DI)+N(HD))/(N(HI)+2N(H2)) instead of HI and DI/HI. However we have 
kept for figure annotations HI and DI/HI.

In Figure 5 we show the TiII/HI ratio as a function of the D/H ratio, 
and error bars associated with the two quantities. D/H error bars are 
taken from Linsky et al. (2006).
It is clear that measurement uncertainties are of the same order of 
magnitude for both related quantities. This will be the case for most 
attempted correlations. Fitting such data using standard or ordinary 
least squares methods can lead to bias in the result.
In order to take into account both x and y errors, we have adopted 
the Orthogonal Distance Regression (ODR) method (also called "total 
least squares method") for our correlative studies. ODR minimizes the 
weighed orthogonal distance from the data to the fitted curve. In our 
case of a simple linear fit of N data pairs ($x_{i},y_{i}$), i=1,N 
with errors $\sigma x_{i}$ and $\sigma y_{i}$ to the function y=ax+b, 
ODR minimizes
  the quadratic sum $\Sigma ((y_{i}-a(x_{i}+dx_{i})-b)/\sigma 
y_{i})^{2}+(dx_{i}/\sigma x_{i})^{2}$, i.e. it adjusts the 
coefficients a and b as well as the departures $dx_{i}$ from the 
initial $x_{i}$ positions.
If the errors on x and y are identical, the summed quantities are 
proportional to the actual orthogonal distances between the data 
points and the fitted curve, hence the name of the method. If one of 
the errors is preponderant, the formula is equivalent to the 
classical least-square method.
We have used the ODRPACK95 package (Boggs et al., 1989) implemented 
in the IGOR 6.0 software to perform our linear fits. In the case of 
linear fits, there is a unique solution to the minimization.
The resulting weighted minimum distances can be visualized in Figure 
5,8,9 and 10.

The ODRPACK95 package works for symmetric errors, i.e. equal positive 
and negative error intervals. If there are substantial differences 
between the positive and negative errors, the data point weight 
depends not only on the distance to the fitted line, but also on 
which side it is located with respect to it, with lower (resp. 
smaller) weight if it is located on the large (resp. small) 
uncertainty side.
In  order to take into account non symmetric errors we devised and we 
use the following iterative method: we first calculate the average 
(mean of positive and negative) error for Xs (resp. Ys) and run the 
ODR linear fit for those mean errors. A simple routine then 
identifies which data points are above or below (resp. to the left or 
to the right of) the fitted straight line and derives from their 
locations which error (the positive or the negative one) is 
appropriate. A second fit is then activated with those new 
appropriate errors. This process is repeated until convergence. For 
linear fits and a relatively small numbers of points, the convergence 
is immediate.
The method is illustrated in Figure 5. We show two data points for 
which the appropriate errors are of opposite signs. For all our 
regressions we show the resulting orthogonal distances between data 
points and the corresponding adjusted locations along the fitted 
line. All quoted uncertainty intervals on the parameters are 
1$\sigma$ and calculated using the error bars from the last fit after 
convergence.

We list in Table 3 two diagnostics for the correlation, the Spearman 
index and the slope interval for the fitted linear relationship. Note 
that the Spearman's rank correlation statistic does not take into 
account error bars, while the error on the fitted parameters does, 
which means that in some cases they
may give different results.

\subsection{Correlations of TiII and FeII abundances with D/H and D/O}

Visual inspection of Figure 5  shows that there is a
positive correlation between TiII/HI and DI/HI, with the noticeable 
exception of the star $\mu$Col. The peculiarity of this target is 
already visible in Figures 4, 6, 7 from Linsky et al. (2006) for FeII 
and SiII. For all three species the abundance is significantly above 
that of the main trend. To a lesser extent a second target also seems 
out of the main relationship, namely LS1274.
  In order to check the influence of the HI column on this result, we 
have displayed the TiII columns as a function of DI in Figure 6.
$\mu$Col is clearly an outlier and LS1274 is also somewhat out of the 
main trend.
We have investigated
  possible reasons for the peculiarity of $\mu$Col, but could not find
anything in the literature. On the other hand, we have seen in the 
previous section that there is strong evidence against an 
interstellar origin of the broad line detected towards this star, on 
the basis of the disagreement between the 3384 and 3242 lines. 
Interestingly, using the narrow line only moves the star into the 
main trend.
  Therefore we performed linear fits to the data, with and without 
$\mu$Col. On the other hand we have chosen to maintain LS1274. Its 
absence does not make any noticeable overall difference.

The Spearman's rank correlation statistic is 0.84 (resp. 0.59) 
without (resp. with) $\mu$Col (15 (resp. 16) points) allowing us to 
reject the null correlation hypothesis at better than 99.97 (resp. 
98) \%.
The parameters for the slope of the linear fit are (1.35 $\pm$0.22) 
and (1.66 $\pm$ 0.29) 10$^{-10}$ respectively (see Table 3). Thus, 
while clearly $\mu$Col modifies the fitted straight line, in both 
cases there is a positive correlation since the most probable 
gradient value is about 6 times the 1$\sigma$ uncertainty on this 
slope.
The reduced $\chi^{2}$ is 1.3 and 1.1 in both cases, thus there is 
possibly some intrinsic variability in addition to the positive 
correlation. The intercept is negative but compatible with zero 
without $\mu$Col.

The results for TiII, keeping the same methods and adding the 4 new 
Ellison et al. targets,
are not noticeably different from those based on our own data
and the Prochaska et al. data only (see Table 3).
However, the presently derived slope is significantly different from 
the one found by Ellison et al. (2007).
There are two reasons for this disagreement. (i) We believe their 
quoted parameters for the linear fit correspond to equal weights for 
all data points because we find the same parameters using their 
listed data and no specific weights. (ii) The fitting method is not 
the same.
Using error bars in X and Y leads to different results. (iii) The 
data sets are not the same.

In order to compare in the most reliable way the TiII results with 
those of FeII, we have also applied the same fitting procedure to 
FeII data, i.e. using the linear iron abundances (and not the 
depletion index) and the ODR method. The data and results are shown 
in Figure 7. It can be seen that the dispersion of iron abundances 
appears significantly greater  than that of titanium. Three targets 
are especially clearly discrepant, with G191-B2B being in the most 
disagreement. Despite this dispersion we have fitted the iron 
abundance vs. D/H distribution without excluding any of the targets.
We will come back to this important point later. Fit parameters are 
listed in Table 3.

Such a positive correlation could be due to random or systematic 
errors in the determination of HI (e.g. H{\'e}brard and Moos, 2003) 
since
HI appears in both related quantities and is by far the most 
difficult measurement. Indeed, changing HI may move the respective 
data point along the straight line fit. This is also the case for 
correlations between metal/O ratios  with D/O ratios. Consequently, 
among our tests we have searched for a correlation between the 
abundance of titanium (TiII/H) and the neutral deuterium to neutral 
oxygen (DI/OI) ratio. This study is restricted to titanium targets 
for which  the OI column is available in the literature.
Interestingly, although the number of data points used is smaller (10 
points), there is clearly a trend, as can be seen in Figure 8. 
Quantitatively, the Spearman test rejects the absence of positive 
correlation at better than 98\%, and the TiII/HI vs. DI/OI slope is 
found to be 6.0$\pm 1.4 10^{10}$ (see Table 3). Such a trend using 
four independent quantities is a definite sign of a positive 
correlation between metals and deuterium.
Adding the two Ellison et al. targets for which D/O is available does 
not significantly change this conclusion. The correlation parameters 
for the 12 point correlation are listed in Table 3.

This result led us to correlate in a similar way the iron (Fe) 
depletion with the DI/OI ratio. Instead of using a logarithmic scale, 
we have used the linear ratio between the abundance of FeII and the 
solar abundance from Asplund et al. (2005). Here the number of 
available measurements is higher (24 lines-of-sight, see Table 2).

Among these three data points there is an extreme 'outlier', G191-B2B. 
The gas towards this target is almost fully ionized, and as discussed 
in section 3.3.1 there may be a strong bias in the evaluation of iron 
depletion because iron is less rapidly ionized than H (see Linsky et 
al., 2006). We have chosen to exclude this target from our 
correlation fit, and we will come back to the ionization problem 
later and justify this choice. This leaves us with 23 sight-lines, 
and the results are shown in Figure 9. Note that G191-B2B
would be completely out of the figure, with a Fe/Fe0 of 0.26. As seen 
with titanium in Figure 7, the data point distribution reveals a 
clear positive correlation. Three points show a rather large 
deviation from the mean curve and seem to have a much higher D/O 
compared to the other targets. Those very distant targets
are indeed the only three targets for which very high D/O ratios have 
been claimed:
WD 1034+001, BD +39 3226 and TD1 32709 (Oliveira et al. 2006).
However, toward those
targets the $\lambda$974\AA\ OI transition has not been used in the 
OI column determination, or was not significantly detected. Not using
this transition could lead to underestimation of N(OI) by factors of 
$\simeq$2, as it has been shown in the cases of Feige 110 
(H{\'e}brard et
al. 2005) and LSE 44 (Friedman et al. 2006). Using this transition,
the OI column density toward BD +39 3226 is also multiplied by a factor of 2.6
(Oliveira et al. 2006). Thus, as suggested by Figure 9, the D/O 
measurements might be overestimated by a factor of 2 to 3 for these 
three lines of sights.

We have thus performed two analyses: firstly using the 3 high DI/OI 
values from the strong line fitting, and secondly after division by a 
factor of 2 of the DI/OI ratios for only these three distant stars, 
hypothesizing that a correction similar to the Feige110 re-evaluation 
is likely to be necessary. When this allowance for correction of line 
saturation effects is included,  the three corresponding data points 
all move back closer to the main stream of data points. We have also 
performed a similar linear fit, excluding the three targets instead 
of revising their DI/OI and found results very similar to those we 
show here. The two fits are displayed in Figure 9 and the linear fit 
parameters before and after correction are given in Table 3. The 
Spearman test rejects the absence of correlation at better than 
99.6\% and 99.99 \% in the first and second case respectively. The 
gradient is somewhat steeper after the correction for the three high 
columns, namely 1.49 $\pm$0.26 instead of 1.28 $\pm$0.22.

The TiII and FeII positive correlations with D/O demonstrate three 
important points. Firstly, potential errors in the measured values of 
HI as the source of a link between D and metals in the gas phase is a 
hypothesis that can now be excluded as the main origin of the 
deuterium-metal correlation. If errors in the measurement of HI  were 
the case, we would simply have randomly distributed points. Secondly, 
the D/O ratio, despite its different evolution as a function of the 
gas column compared to DI/HI, is shown to vary in a specific way, and 
shows a link to the metal abundance. Third, D/O cannot be linked to 
metal depletion principally through astration effects, since this 
would produce the opposite trend, with D/O being anti-correlated with 
the degree of astration and thus with metal abundance.

\scriptsize
\begin{table*}
\caption{Correlation parameters. LHW= this paper, PTH05 = Prochaska 
et al. (2005), EPL07= Ellison et al. (2007), Lin06 = Linsky et al. 
(2006), Ol06 = Oliveira et al., (2006)}
\label{table3}
\centering
\begin{tabular}{c c c c c c c}     
\hline\hline
Correlation  & Targets & Data points& a$\pm1\sigma$ & b$\pm1\sigma$ & 
Spearman test & $\chi^{2}$/N\\
y=a+bx & x= list intersection & & & correlation  & &\\
\hline

(TiII/HI) vs (DI/HI) & LHW + PTH05 & 16 & (-3.6 $\pm$ 2.9) 10$^{-10}$ 
& (1.66 $\pm$ 0.29) 10$^{-4}$   & 98\% & 1.3 \\
(TiII/HI) vs (DI/HI) & Same without $\mu$Col  & 15 & (-0.8 $\pm$ 2.3) 
10$^{-10}$  &  (1.35 $\pm$ 0.22) 10$^{-4}$  & 99.97\% & 1.1\\
(TiII/HI) vs (DI/HI) & Same + EPL07 (no $\mu$Col)& 19 & (-0.6 $\pm$ 
2.32) 10$^{-10}$  & (1.29 $\pm$ 0.21) 10$^{-4}$    & 99.5\% &1.0\\
(FeII/HI) vs (DI/HI) & Table 2 Lin06 & 38  & (-1.56 $\pm$0.76) 
10$^{-2}$  & (4.96  $\pm$ 0.65) 10$^{+3}$  & 99.7\% &1.6\\
(TiII/HI) vs (DI/OI) & (LHW + PTHO5) x (Oli06) & 10 & (-1.24 $\pm$ 
3.29) 10$^{-10}$ & (6.05 $\pm$ 1.45) 10$^{-8}$    & 98\% &2.7\\
(TiII/HI) vs (DI/OI) & Same+EPLO7 & 12 & (-2.07 $\pm$ 3.46) 
10$^{-10}$ & (6.46 $\pm$ 1.50) 10$^{-8}$    & 97\% & 2.3 \\
(Fe/Fe0) vs (DI/OI) & (Table 2 Lin06) x (Oli06) & 23  & (3.5 $\pm$ 
5.6) 10$^{-3}$  &  1.28 $\pm$ 0.22  & 99.6\% & 1.3 \\
(Fe/Fe0) vs (DI/OI) & Same with 3 D/O \%2 (see text) & 23 & (-0.5 
$\pm$ 6.6) 10$^{-3}$  & 1.49 $\pm$ 0.26   & 99.95\% & 0.7\\
(TiII/DI) vs (LogNH) & LHW + PTH05 (no $\mu$Col) & 15 & (5.9 $\pm$ 
2.5) 10$^{-4}$ & (-2.34 $\pm$ 1.24) 10$^{-5}$   & 25\% & 1.46\\
(TiII/DI) vs (LogNH) & same + EPL07 & 19 & (7.6 $\pm$ 1.8) 10$^{-4}$ 
& (-3.2 $\pm$ 0.9) 10$^{-5}$   & 64\% & 1.8 \\

\hline\hline
\end{tabular}
\end{table*}
\normalsize

\subsection{Gradient comparisons}
\subsubsection{Comparisons between titanium and iron}

It is interesting to compare the variation amplitude of titanium and iron.
Titanium is strongly refractory and one expects a higher amplitude of 
the gas-phase abundance variation, as argued by Ellison et al. 
(2007). Figure 5 shows the FeII vs. DI/HI gradient from Figure 7 
superimposed on the TiII vs. DI/HI gradient. Contrary to expectations 
based on condensation temperatures, TiII varies slightly less than 
FeII, although as shown in Table 3 and taking into account the two 
error bars, the two gradient intervals nearly overlap. In the case of 
DI/OI the two gradients are compatible, but the TiII vs. DI/OI 
gradient is less precisely defined than for DI/HI.

We believe that the absence of a stronger gradient for TiII is not an 
inevitable concern with respect to the deuterium depletion hypothesis 
of Linsky et al (2006) and that the observed superiority or 
similarity of the iron slope can be explained by ionization effects. 
Linsky et al. (2006) have already discussed the potential impact of 
differential ionization of FeII and HI on the FeII/DI ratio 
variation. The ionization potential of FeII is 16.2 eV, and as a 
consequence hydrogen may ecome ionized by the radiation field while 
FeII remains only slightly affected.
The case of G191-B2B , the extreme outlier, is typical. From their 
analysis of high resolution HST-GHRS spectra, Lemoine et al (1999) 
inferred the existence of two main clouds along the LOS to the hot 
white dwarf. For the main HI cloud (B) the FeII/HI ratio is 0.16 
10$^{-5}$ while for the second, strongly ionized cloud (A), the same 
ratio is 2.25 10$^{-5}$, i.e. 13-14 times above cloud B. The 
existence of the ionized fraction contained in cloud A has the effect 
of increasing the overall ratio (for A+B) by a factor of more than 3 
compared to cloud A alone. This is a very strong effect, and it is 
clear from Figure 7 that lines-of-sight filled with significantly 
ionized gas,  such as G191-B2B or 36 Oph have a systematically higher 
singly ionized iron abundance, that probably does not reflect the 
actual iron abundance.
As a consequence it is likely that a high ionization degree, a 
phenomenon expected to be present in shock- and photo-heated regions, 
i.e. in regions with the strongest evaporation and deuterium release 
from grains, also tends to increase the FeII/HI ratio. In other 
words, in the absence of precise corrections for the gas ionized 
fraction, iron release in the gaseous phase and apparent abundance 
increase due to ionization balance effects act in a cumulative way. 
This tends to increase the FeII/H ratio and the larger gradient for 
FeII may result from this trend.
As we explained earlier G191-B2B is an extreme case and we have 
removed it from the correlation with DI/OI shown in Figure 9, which 
has the effect of decreasing the slope. Probably other remaining 
targets such as WD2211-495 have an effect on the slope of the linear 
fit. G191B2B is also the most discrepant point for the FeII/HI vs. 
DI/HI correlation, but because there are two times more targets for 
this correlation it is not isolated, and this is why we have kept it. 
Looking at Figure 7 it is clear that three data points including 
this star tend to increase the gradient of the correlation. These 
ionization effects are absent in the case of TiII, due to the 
similarity of the potentials. This is why we believe that the absence 
of the expected stronger gradient for titanium compared to iron is 
due to an overestimate of the latter.

\begin{figure}
    \centering
   \includegraphics[width=8cm]{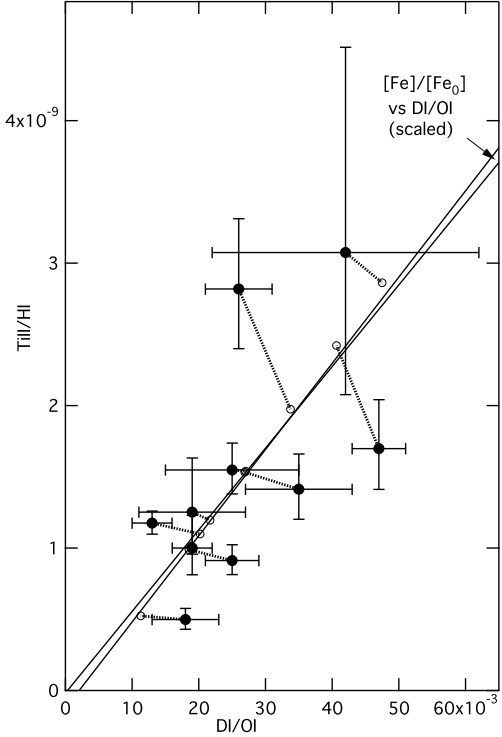}
       \caption{Left: Ionized titanium vs. DI/OI. The corresponding 
linear fit for FeII (Figure 5) is shown for comparison.}
          \label{lintih_do}
\end{figure}

\begin{figure*}
    \centering
   \includegraphics[width=14cm]{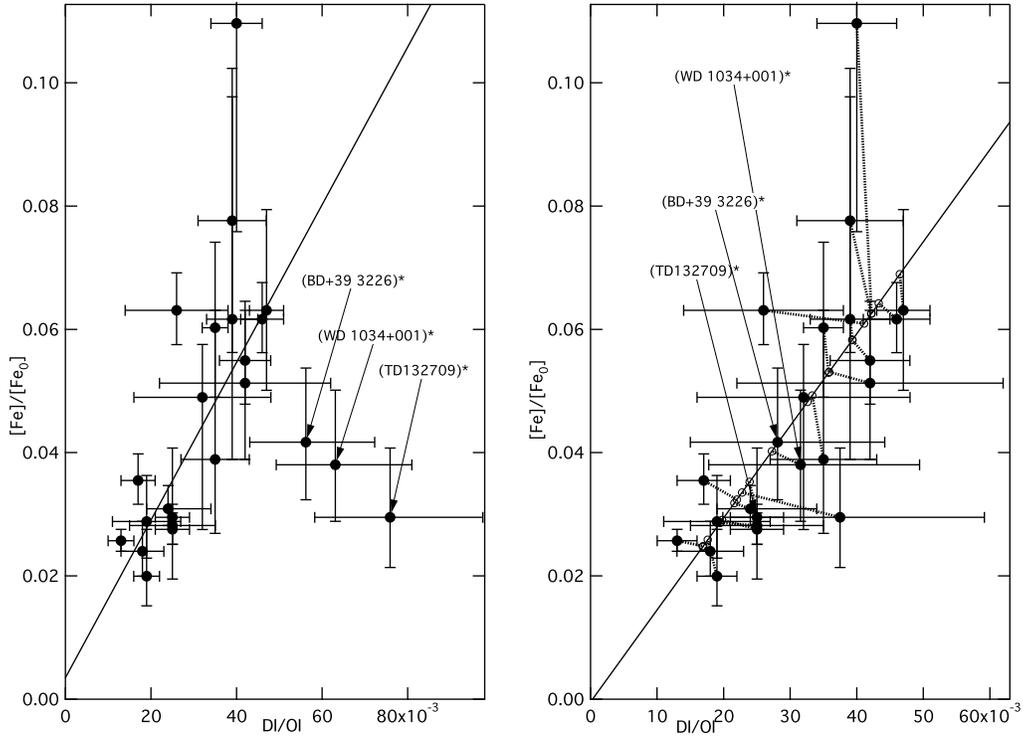}
       \caption{Iron gaseous abundance (normalised to the Asplund et 
al abundance) and Deuterium to Oxygen ratio. a: Published DI/OI 
results. b: After division by a factor of 2 of the three DI/OI ratios 
for the three distant stars (see text).}
          \label{FigFevsDO}
\end{figure*}

\begin{figure*}
\centering
\includegraphics[width=13cm]{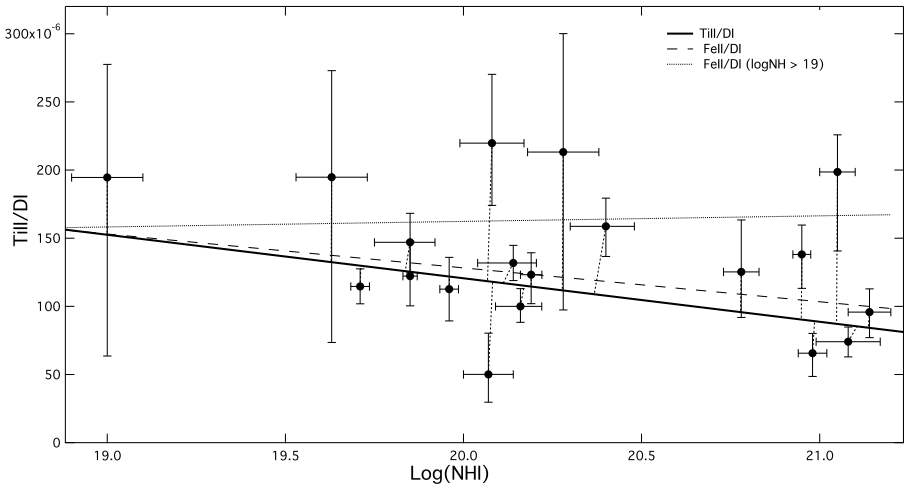}
\caption{TiII/DI ratio as a function of the HI column: all available 
measurements. The linear fit using the Orthogonal Distance Regression 
method is shown as a solid line. The slopes found by Linsky et al. 
(2006) for the FeII/DI ratio are shown as dashed and dotted lines 
(resp. all data, and data with logN(HI) $\geq$ 10$^{19}$ cm$^{-2}$).}
\label{tidvslognh}
\end{figure*}

\subsubsection{Comparison between the titanium and iron gradients vs. 
D/H and vs. D/O}

While the positive correlation between the metal abundance and D/O 
can not be explained by astration (it would be a negative trend), it 
may be well explained, as for D/H (Linsky et al., 2006), by the 
depletion hypothesis. Here we compare the two gradients obtained from 
the previous correlations of TiII and FeII with D/H and D/O.

Along with deuterium and metals that evaporate from grains and are 
released in the gaseous phase, it is reasonable to expect some 
simultaneous release of oxygen. However the relative increase of the 
OI concentration due to this release is much smaller than for DI, 
because the oxygen abundance is the gaseous phase is much higher than 
for deuterium. Also, data show a quasi-constancy of the OI/HI ratio, 
showing that an increase due to dust evaporation is negligible.
Therefore, in the depletion scenario, for a given increase of the 
metal gaseous abundance, i.e., of dust evaporation, DI/OI should 
increase in the same proportions as DI/HI.
Figures 5,7,8, and 9 and Table 3 show that this is the case globally, 
i.e. the metal abundance, the D/H ratio and the D/O ratio all vary by 
a factor of 3-3.5, and there are significant linear correlations 
between metals and D/H on one hand, metals and D/O on the other. 
However, there is no simple proportionality between the three 
quantities, since some lines of sight with very high D/H ratios do 
not present correspondingly high D/O values (H\'ebrard et al, 2005).

It is striking that error intervals for the fit parameters allow (or 
nearly allow) a simple proportionality between Ti or Fe and D ( the 
linear fit in this case goes through zero). This implies that D 
increases as rapidly as metals. In the framework of the deuterium 
depletion scenario of Draine (2004) and Linsky et al. (2006), one 
would expect a preferential release of deuterium from PAHs and the 
carbonaceous envelopes of grains, whereas metals (O and Fe) are 
instead preferentially removed from the cores of interstellar grains, 
and therefore a strict proportionality is somewhat surprising.

The comparison between the residuals for TiII/HI vs. D/H 
($\chi^{2}$/N = 1.3 and 1.1, see Table 3) and D/O ($\chi^{2}$/N = 2.7 
and 2.3) shows that the fit is of significantly better quality vs. 
D/H than vs. D/O. We believe this is due to the fact that in the 
former case N(H) appears in the two correlated quantities. As a 
consequence, measurement errors on H tend to produce dispersion along 
the fitted line, that does not increase the $\chi^{2}$. This shows 
that a fraction of the correlation between TiII/HI and D/H is due to 
uncertainties on HI columns. In the latter case of the correlation 
with D/O, measurement errors in H bring a dispersion along the 
ordinate axis that appear in the $\chi^{2}$.
Such effects for iron are not seen because the residuals are 
dominated by the ionisation biases that introduce a strong 
dispersion. Especially, we have removed G191B2B for the correlation 
with D/O and kept it for D/H, which explains the higher $\chi^{2}$/N 
for D/H.

\subsubsection{Correlation between titanium abundances and column-densities}

We have searched for deviations from proportionality and for a 
potential link between the titanium to deuterium abundance ratio and 
the HI column. Linsky et al.(2006)
have showed evidence for an increase of the DI/FeII ratio along with 
the HI column. More precisely, there seems to be an increase from 
very low columns (10$^{18}$ cm$^{-2}$) to columns of the order of 
10$^{19-19.5}$ cm$^{-2}$ and constancy above log(NH) $\simeq$ 19.5 
cm$^{-2}$. They interpreted this trend as evidence for a quasi 
proportionality of D and Fe abundances in general, and attributed the 
low to medium column increase as being
due to significant ionisation of the diffuse gas close to the Sun, an 
effect similar to what we have invoked for the Ti vs. Fe comparison. 
As discussed by the authors and previously in this paper for HI, in 
ionized gas D may be more easily ionised than FeII, leading to an 
overestimate of the abundance of FeII at low columns within the Local 
Bubble.

  In the case of TiII, ionisation effects are not supposed to have any 
contribution due to the coincidence between the two ionisation 
potentials of TiII and HI (or DI). It is therefore particularly 
interesting to perform the same correlation test as the one done by 
Linsky et al. but for TiII instead of FeII. Figure 10 shows the 
TiII/DI ratio as a function of logN(HI). Similarly to the Linsky et 
al. (2006) results for FeII, the smallest HI columns seem to 
correspond to the highest TiII /DI ratio.
While the trend is not as clearly defined as in the correlation 
analyses in the previous sections, with a Spearman rejection of null 
correlation at about 75\% probability only (and a negative result if 
we do not include the high column targets of Ellison et al (2007), 
see Table 3), the least-squares linear fit with the ODR method taking 
into account error intervals on TII/DI and logN(HI)  demonstrates a 
negative slope at better than 3.5 $\sigma$s. However, the slope is 
very strongly dependent on the three points with high columns and 
small error bars and more or better data are needed to assess this 
result.

The sign of this inferred TiII/DI gradient cannot be explained by 
less astration
for high columns, i.e. far from the Sun, since it would contradict 
the increase of OI/HI  at large columns found by Andre et al. (2003) 
and more recently by Oliveira et al. (2006).
Tentatively, the sign of the slope may correspond to preferential 
evaporation of deuterium compared to titanium at the highest columns, 
i.e. in very dense and colder regions which become preponderant along 
very distant lines-of-sight. In other words, D in this case is 
released into the ISM before metals, and only in warmer and more 
tenuous media are the release rates identical.

Superimposed on the Tii/DI vs. logN(HI) fitted relationship are shown 
the corresponding results of Linsky et al. (2006) for FeII. Ordinates 
have been multiplied by a constant, and we are interested here in the 
slope only. Our results that correspond to logN(HI) $\geq$ 19 should 
be compared to the flat curve calculated by these authors for targets 
with logN(HI) $\geq$ 19.3.

As we already mentioned, titanium data are very appropriate here 
because of the absence of ionization effects that may introduce 
biases. However there is a lack of measurements with  logN(H) smaller 
than 10$^{19}$ cm$^{-2}$ to confirm unambiguously a trend.

\section{Discussion and Conclusions}

We have observed nine southern hemisphere D/H target stars at high spectral 
resolution and derived the column-densities of interstellar ionised 
titanium
towards those targets. A comparison between the D/H ratio and the 
titanium depletion reveals a clear correlation and reinforces
the result of Prochaska et al. (2005) and the more recent result of 
Ellison et al. (2007).
Our work extends the correlation over a broader range of interstellar 
HI columns, namely more than two orders of magnitude, showing that 
the trend extends down to log N(HI)=19.0 cm$^{-2}$. We also make use 
of the orthogonal distance regression method for our linear fits, in 
order to take into account errors in quantities both in abscissa and 
ordinate.

One star is far out of the correlation, $\mu$Col. This target is 
already strongly discrepant in correlations shown by Linsky et al 
(2006) between D/H and iron and silicon abundances. We have shown 
evidence here for a non-interstellar origin of the broadest 
absorption feature and note that excluding this absorption would 
bring the star back to the main trend.

  We have compared both the Ti and Fe depletions with the DI/HI ratio, 
but also with the DI/OI ratio, when available. For both metals there 
is a clear relationship also with DI/OI, which demonstrates that the 
metal abundance- D/H correlation is not primarily due to 
uncertainties in HI column measurements, and that D/O is also 
correlated with metal depletions in the ISM.

The ionized titanium correlation gradient is found to be similar to 
or slightly smaller than the corresponding gradient for ionized iron. 
Ellison et al (2007) have emphasized that given its high condensation 
temperature one would expect the titanium gradient to be the highest. 
We have argued here that ionization effects do affect FeII/HI and 
result in an overestimate of the gradient, while the titanium 
abundance is unaffected due to the nearly exact coincidence between 
TiII ad HI ionization potentials. Thus we believe that the slopes 
measured for both metals are not a concern with respect to the 
depletion hypothesis as the source of deuterium abundance variations. 
Further work is needed to correct for the ionization and obtain a 
more reliable comparison between the different metal-deuterium 
relationships.

There is some (although weak) evidence that the TiII/DI ratio 
decreases towards high column densities. This may signify that D is 
preferentially released into the ISM (when compared to titanium) in 
the dense regions that are predominant along distant lines-of-sight. 
It is important to investigate further this potential relationship, 
because, if confirmed, it would be a first indication of the 
different sources of interstellar D and metals. Recording more 
titanium data along shorter lines-of-sight is essential. TiII, that 
does not suffer ionisation biases, is an ideal tracer here.

It remains to explain why very high D/H ratios do not have 
corresponding very high D/Os, as already argued by  H{\'e}brard et al 
(2005).  This is reinforced here by our findings (section 3.2 and 
Figure 9a) that the three high D/O values found by Oliveira et al 
(2006) may be overestimated.
   Additionally, Oliveira and H{\'e}brard (2006) derive a high D/H for 
a very distant sightline while one would expect in this case a very 
strong depletion.
Certainly some of the physical processes are not yet understood.
On the other hand, since very discrepant points appear in the 
correlation with D/H, but none with D/O, the high D/H ratios may be a 
consequence of underestimated H columns.

Assuming proportionality between metals and D depletion, one can derive
a mean O/H ratio from the ratio of the Ti/H vs. D/H and Ti/H vs. D/O 
slopes (or similarly for Fe). Using TiII one finds (Table 3) 6.05 
10$^{-8}$/1.35 10$^{-4}$ = 448 ppm using our targets and the 
Prochaska et al. targets, and 6.46 10$^{-8}$/1.29 10$^{-4}$ = 501 ppm 
using all targets.  Using FeII one finds 1.49/4.96 10$^{3}$= 300 ppm. 
There must be some significance to this difference that remains to be 
elucidated. These two values are respectively higher and lower than 
the mean value of 347 ppm derived by Cartledge et al. (2004) within 
800 parsecs.
It is also interesting to compare with the range of solar 
photospheric O/H values recently proposed by Grevesse et al. (2007). 
The value derived here from the TiII data, 480-500 ppm, is within 
their quoted range of 407-513 ppm, while the value derived from FeII, 
300 ppm, is significantly below.
These calculations suggest that the slopes derived from the linear 
fits using TiII are meaningful.

Is there something to learn about the local ISM history from the D/H 
abundance pattern? The large-scale event that has given rise to the 
Gould belt of enhanced star formation is about 60 Myrs old (see e.g. 
Perrot \& Grenier, 2003). Two main classes of scenarios have been 
proposed for the belt origin: the interaction between a giant 
external cloud and the disk on one hand (Comeron and Tora, 1994, 
Olano, 2001) and a strong explosive event on the other (Olano, 1982). 
Because the age of the belt is smaller than the mixing time
of gases (a few hundreds Myrs, see e.g. simulations by De Avillez and 
Breitschwerdt, 2004), in the former case of cloud collision the belt 
may have left imprints in the form of incomplete mixing between disk 
and "fresh" gas.  In the latter case there must be dust evaporation 
at the shocked front regions of the belt but no gas mixing. A link 
between the Gould belt expansion and the D/H distribution has been 
suggested because highly variable D/H regions shown by Linsky et al 
(2006, Fig. 1) are located at distances (inferred from the H columns) 
that roughly correspond to the front of the expanding Gould belt of 
young stars and supernovae (Lallement, 2007). Whatever the origin of 
the belt, the subsequent and observed expansion produces star 
formation and supernovae at the periphery of the belt, i.e. 
evaporation of dust in the shocked and heated regions of its 
periphery, which could explain the observed variability of D/H at 
intermediated columns (10$^{19.5-20.5}$ cm$^{-2}$). The difference 
between the two scenarios for the belt origin is that in the cloud 
collision case some deuterium abundance variability due to gas mixing 
must add to abundance variation due to grain destruction, and there 
must be a larger amount of less astrated gas within the Gould belt as 
compared with outside the belt.
There are no signs of astration variations that follow from the 
present correlative analysis, in agreement with the constancy of O/H 
up to large distances (Andr{\'e} et al., 2003). If D and metal 
abundance variability has some link with the belt, then the absence 
of signs of differential astration at columns lower than 10$^{21}$ 
cm$^{-2}$ seems to favor the explosive event scenario.

Independantly of a potential effect of the Gould belt on the D/H 
distribution, our results provide a support for the ideas, originally 
suggested by Draine et al. (2004) and Linsky et al. (2006), that 
time-dependent deuterium depletion onto dust grains is responsible 
for the variability of the D/H ratio, and for the low values of D/H 
observed for many lines of sight beyond the Local Bubble. On the 
other hand, it remains to explain why very high D/H ratios do not 
correspondingly present very high D/O values. This is important for 
the precise determination of the present galactic D/H ratio.

\begin{acknowledgements}
  We thank our referee Jeffrey Linsky for very carefully reading of 
the manuscript. His remarks and criticisms resulted in a strong 
improvement of the paper.\\
  We thank Francis Dalaudier from Service d'A{\'e}ronomie for his 
useful recommendations on the fitting procedures and interesting 
suggestions for non symmetric error treatments, and
Alain Lecavelier des Etangs from IAP for his support of the observing 
proposal and useful discussions.
\end{acknowledgements}

\end{document}